# Synchronous Relaying Of Sensor Data

Rahul R Upadhyay
Mechanical Engineering Department, BTech(Final Year)
BBDNITM, Lucknow, India
rahulrupadhyay91@gmail.com

**Abstract**

*In this paper we have put forth a novel methodology to relay data obtained by inbuilt sensors of smart phones in real time to remote database followed by fetching of this data . Smart phones are becoming very common and they are laced with a number of sensors that can not only be used in native applications but can also be sent to external nodes to be used by third parties for application and service development.*

**Keywords**

PhoneGap, Ajax, UDP/TCP, Cordova.

## 1. Introduction.

The wide range of sensors that are present in smart phones can be used not only for making applications for the smart phone but can also be relayed to external nodes which can utilize this sensor data for making a pool of applications and services.

In this paper, we describe a novel method to fetch and transmit the values of sensors in a smart phone to a remotely hosted DAT[25] file. The values thus stored can be fetched by a number of DRN[23], or data retrieval nodes and can be utilized as decision attributes in various applications ranging from mechatronics to solely software based.It consists of three nodes, data transmittal node which is also called local node, web-host node and DRN or data retrieval node.

In this paper we are demonstrating our method on Android platform and we relay accelerometer sensor values. DTN or local node is basically a hybrid app developed using PhoneGap that forms a gap b/w the native functionalities of the smart phone and the web platform. It fetches the sensor data and relays it every 350 ms to a PHP file using GET method via AJAX through synchronous channel.Web host node is basically a PHP script that fetches the accelerometer values and puts it in DAT file with a delimiter. [25] [26]

Whereas the DRN or data retrieval node fetches the values in the DAT file and segregates the string data using the delimiter and converts it into a numeric data type before using these values in whatsoever way.

## 2. Related Work.

Applications have been developed that can send sensor and location information of a smart phone device as Comma Separated Values or CSV lines using UDP or user datagram protocol. [18][13]

The methodology mentioned in this paper in not an individual technology but rather combines and integrates a number of technologies including hybrid smart phone app development, web based technologies like PHP and real time data fetching on network. VANET is a similar technology that stands for Vehicular AdHoc Networks and integrates several technologies including GPRS, GPS etc in order to obtain an intelligent transportation system. The communication between vehicles is achieved using a dedicated short range communication (DSRC)[1]. The packets of data are transmitted via TCP / IP or UDP.

Smart phone applications have been developed that relay sensor data using TCP [17]. TCPs are basically COT or connection oriented transmission and there is acknowledgement and retransmission facilities. Although TCPs are considered "disciplined" transmission channels, they are comparatively slow and thus UDPs are preferred over them. [15] [21]

UDPs are widely used to transmit sensor data from smart phones owing to its higher speed. In UDP, no prior communication channel is setup before relaying data. UDPs are considered as unreliable as some data packets get lost in communication channel or come out in haphazard fashion in the information stack. For this reason TCP is used in places where the sequence of information is primary need and speed the secondary.[9]

# 3. Method Overview.

## 3.1 About Sensors

Nowadays smart-phones are laced with a wide range of sensors that measure motion of the smart-phone device, proximity of nearby objects to the device, orientation of the device, direction of the device etc. Some devices are even laced with temperature and gravity sensors. [16] [14]

The native applications of the device use the data obtained using these sensor devices to take decisions. For instance, depending upon the distance between the ear and the receiver of the smart-phone, measured by the proximity sensor of the phone, the screen is activated or blanked during phone calls.[26]

Like any sensor, the huge cache of sensors that are inbuilt in a smart-phone can also be utilized for external purposes and can be used for developing a massive range of applications and services in several sectors owing to the wide increasing reach of smart phone market. For instance, GPS location soft-wares use the inbuilt GPS sensors to track the movement of the phone and can be used in human tracking and security. [22]

ANDROID supports three major classes of sensors, which includes

- Motion Sensors
- Environmental Sensors.
- Position Sensors

In this paper we will be demonstrating our methodology using motion sensors strictly. Motion sensors measure acceleration forces, rotational forces in three mutually perpendicular directions .This category of sensors include, RVS which stands for rotational vector sensors, gyroscopes, gravity sensors and also accelerometers. [20]

## 3.2 Accelerometer

Accelerometer is a motion sensor that detects the change of orientation and the position of the smartphone device with respect to the orientation and location of the device in the current situation.

Accelerometer, detects change in 3D movement of the smartphone device along X, Y and Z mutually perpendicular direction. In this paper we will be relaying the values of obtained by the android accelerometer to our DAT file from where it will be fetched by our control software that is remotely kept. [23]

## 3.3 Methodology

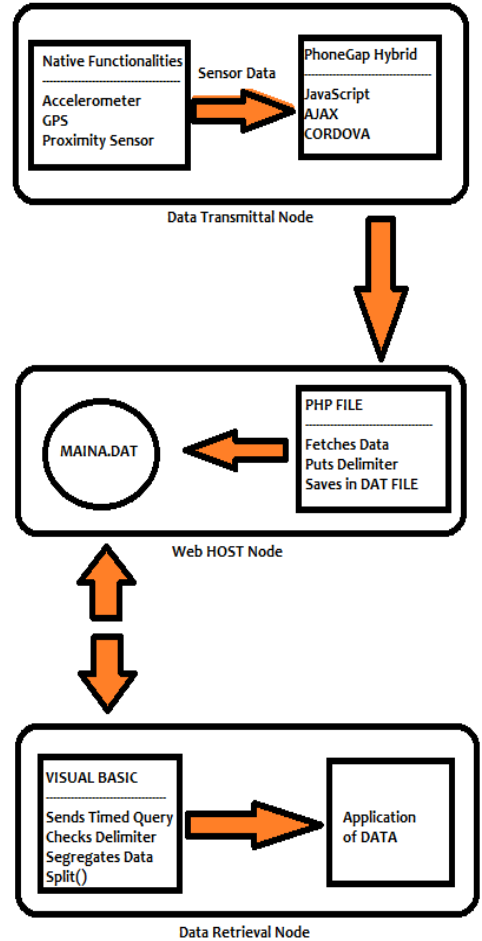

Figure 1. Flowchart of Entire Methodology.

As shown by the flowchart, sensor data of the smart-phone is fetched by the PhoneGap hybrid app which uses JavaScript and Cordova, that bridges the gap between the native capabilities of the device and web based technologies.

The obtained data (in this case, accelerometer values) is relayed every 350 microseconds to a php script using GET method. This node of the method is also called WebHost node. The php script puts this set of data in a sequence using a suitable delimiter in a DAT file. [14]

The data retrieval node fetches the data and every 100 microseconds and segregates the accelerometer values in x, y and z axis and converts them into a numeric datatype for further use. [11]

## 4. Data Relay Node.

DRN or Data relay node is basically an application installed on the smartphone device through which sensor data has to be relayed to the WBN or web host node. DRN in its most basic form fetches the sensor data on a timely basis, x microseconds and transmits this data synchronously to the WBN using AJAX. The value of x is chosen during programming phase, however the DRN apk (for android devices) can be edited so as to add an extra flexibility of altering the transmittal frequency and choosing relaying method between synchronous and asynchronous during runtime. [4] [8]

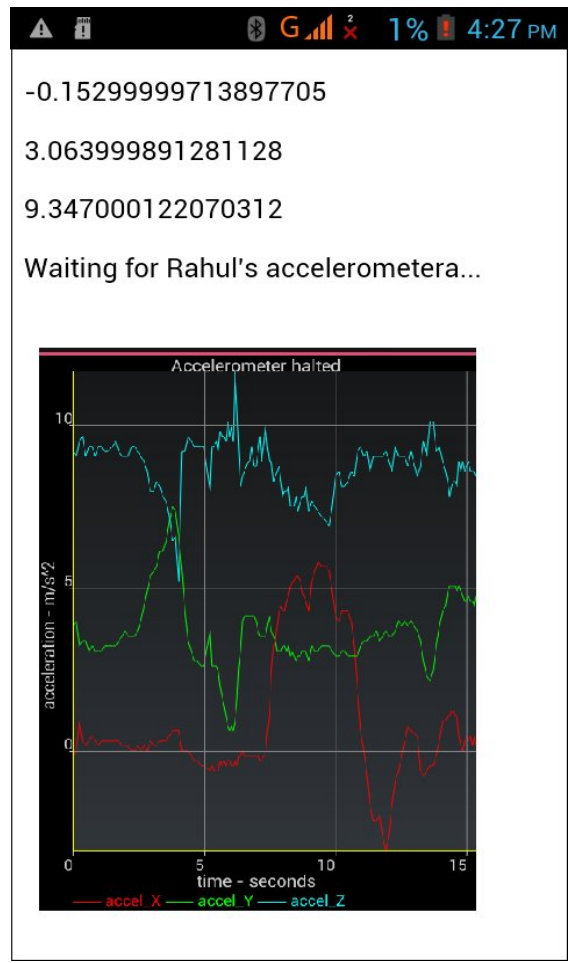

Figure 2. Snapshot of the DRN apk on Android.

Figure1 shows the snapshot of the Data relay node in ANDROID platform. The first number signifies change in the orientation about the X axis followed by Y and Z. This application fetches the accelerometer data followed by relaying the data to the cloud along with printing the data on the screen for viewing. [10]

## 4.1 DRN Hybrid (Apache Cordova)

DRN is developed using PhoneGap technology. It is basically a MDF or mobile development framework through which mobile device applications can be built without using device specific languages like Objective-C but rather web based languages like JavaScript, HTML5 and CSS. PhoneGap forms a programming bridge to use JavaScript to actuate and use native functionalities of the device.

The applications developed using PhoneGap are hybrid in nature which means that the applications are not completely native and neither completely web based. This is owing to the fact that the applications neither use the device's own User-Interface framework and neither purely web based as they are packaged as apps and have access to the Application Programming Interface (API) of the native device. [23][11]

## 4.2 GET Synchronous Relay

In AJAX, XMLHttpRequest communication channel can be either synchronous or asynchronous. In asynchronous request the block of code will begin its execution and the next statement will be called irrespective of the fact whether the previous code has been executed or not. In synchronous request, the block code will begin execution and the control would not transfer to the next statement until that present code has finished its execution[3]. In general, asynchronous request is preferred and this methodology also used it to relay data initially. [24]

We realized that there were instances when although the accelerometer data was continuously updating on the device screen, the web based DAT file (storing the data via PHP) malfunctioned owing to massive traffic of requests. This was owing to the fact that the data was being fetched and relayed every 350 microseconds and sometimes during the course of time that the XMLHttpRequest was in action, a second or third cycle of request was also executed[5]. The results were far better when we used synchronous relaying method as the control of program did not transfer to the second statement until the XMLHttpRequest was executed.

**xmlhttp.open("GET",str,false); xmlhttp.send();**

For sending a request to the PHP code which stores the sent accelerometer values to the DAT file, GET method is used as it is faster than POST method. Here "str" stores the current accelerometer values.

### 4.3 JavaScipt Code

```
<!DOCTYPE html>
<html>
 <head>
  <title>Acceleration Examplea</title>
  <script type="text/javascript" src="cordova.js"></script>
  <script type="text/javascript" src="js/index.js"></script>
  <script type="text/javascript" charset="utf-8">
  var watchID = null;

  document.addEventListener("deviceready", onDeviceReady, false);

  // Wait for PhoneGap to load
  // PhoneGap is ready
```

Cordova consists of native and JavaScript code bases. Usually native code takes less time to open with respect to JavaScript code . JS code is loaded once DOM is loaded hence[2][6] it is necessary that no action is executed until both are loaded and ready to use. "deviceready" function ensures that the application starts when Cordova has completely loaded.

```
function onDeviceReady()
{
   startWatch();
}
// Control transferred to startWatch when PhoneGap is ready

function startWatch() {

   // Update acceleration every 3.5 seconds
   var options = { frequency:350 };

   watchID = navigator.accelerometer.watchAcceleration(onSuccess, onError, options);
}
```

Once Cordova is loaded, the control is transferred to the startWatch() function which calls the on Success function every 350 microseconds .

```
var xmlhttp;
xmlhttp=new XMLHttpRequest();

function onSuccess(acceleration) {
   x=document.getElementById("demo");   // Find the element
   x.innerHTML=acceleration.x;
   y=document.getElementById("demo1");  // Find the element
   y.innerHTML=acceleration.y;
   z=document.getElementById("demo2");  // Find the element
   z.innerHTML= acceleration.z;

   var ax=acceleration.x;
   var ay=acceleration.y;
   var az=acceleration.z;
   var str="../man.php?Editbox1="+ax+"&Editbox2="+ay+"&Editbox3="+az+"&Button1=Submit";
   xmlhttp.open("GET",str,false); //synchronous
   xmlhttp.send();
}
```

OnSuccess () function fetches the current orientation and position value of the device using accelerometer sensor each time it is called. Besides this, the values are sent to the PHP code using GET method via AJAX – synchronous channel. The data is also printed on screen of the device. [23]

```
</script>
</head>
<body>
<p id="demo">X Axis</p>
<p id="demo1">Y Axis.</p>
<p id="demo2">Z Axis.</p>
  <div id="accelerometer">Waiting for Rahul's accelerometera...</div>
</body>
</html>
```

### 5. Web Host Node.

Web Host Node or WBN is basically a PHP code to which accelerometer data is sent using GET method. This PHP code accepts the accelerometer values in three mutually perpendicular direction and stores it in the DAT file by putting a "<br>" delimiter . Using this delimiter the values of x, y and z axis are segregated in the DRN or Data Retrieval Node. [24] [23]

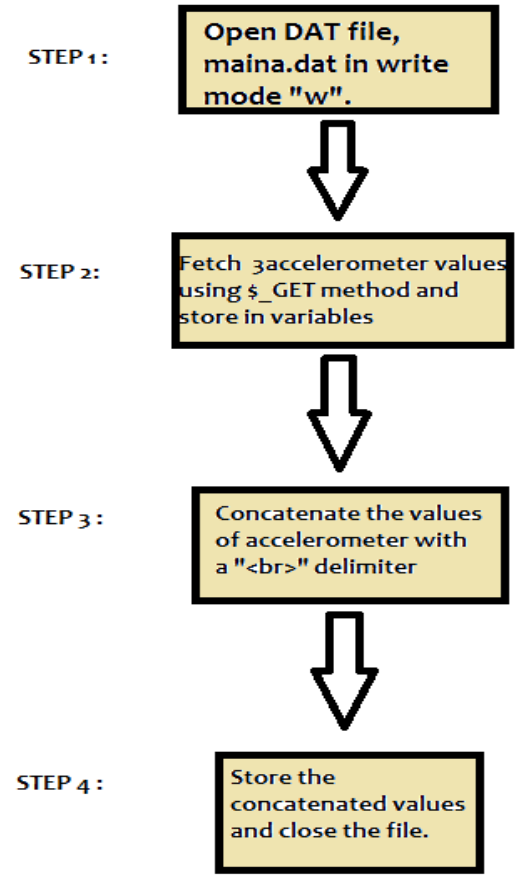

Figure 3. Flowchart of The PHP algorithm.

### 5.1 PHP code

```
1  <?php
2
3  if(isset($_GET['Button1']))
4  {
5     $myFile = "maina.dat";
6     $fh = fopen($myFile, 'w') or die("can't open file");
7     $stringData = $_GET['Editbox1'];
8     $stringData1 = $_GET['Editbox2'];
9     $stringData2 = $_GET['Editbox3'];
10    $stringDataa=$stringData;//"<br>".$stringData1."<br>".$stringData2;
11    fwrite($fh, $stringDataa);
12    fclose($fh);
13 }
14
15 ?>
```

## 6. DRN or Data Retrieval Node.

Data Retrieval Node or DRN is any piece of code that fetches the values in the DAT file. In this paper we are using Visual Basic to fetch the data from the remote webhost node followed by segregating the data depending upon the delimiter and then printing the data. [25]

It should be noted that the data is obtained in the form of a string and after segregating the data, it can be converted into a numeric database if it has to be used for any application or calculation purpose.
In the data transmittal stage we relayed the data every 350 micro seconds so it is necessary that we fetch the data within a time frame less than 350 ms. Here we have chosen 100 micro seconds between each Inet request to the data file. [7]

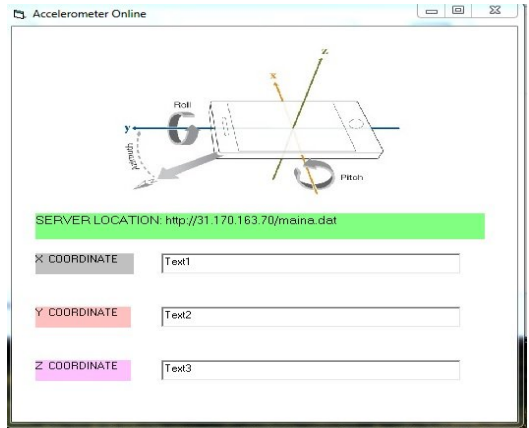

Figure 4. Snap shot of DRN.

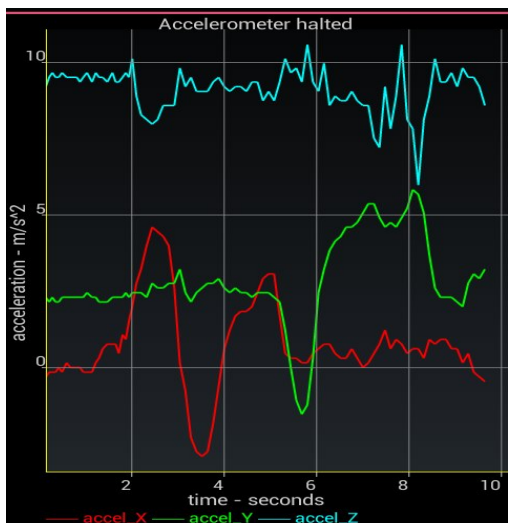

Figure 5. Accelerometer Data( Snapshot from *Accelerometer Kinetics*).

## 6.1 Visual Basic Code

```
Private Function add()
Dim contentfile As String

contentfile1 = Inet1.OpenURL("http://hellow21.comoj.com/maina.dat")
con = Split(contentfile1, "<br>")

Text1.Text = con(0)
Text2.Text = con(1)
Text3.Text = con(2)

End Function

Private Sub Timer1_Timer()
Dim val As String
an = add()
End Sub
```

Inet.OpenURL function is used to fetch the contents of the maina.dat on remote server . The obtained value is segregated using split function with delimiter "<br>".
A timer function set with 100 ms calls the main function.

## 7. Conclusion and Applications

Several methods have been proposed wherein the data is sent as UDP packets on private IPs. But using this methodology data can be sent on public ip's in the simplest possible way without socket handling and firewall restrictions. It uses simple AJAX and PHP channels and data is sent as raw code rather than in packets. [23] [19]

This method can be used to relay sensor information such as GPS location, compass direction and orientation to make applications and services for day to day life. For instance, relay of GPS location of a device can be fetched by DRN set up in a car door which would open it , once the device is nearby.

## 8. Future Scope and Improvement

The DTN application is developed using PhoneGap technology which render it of hybrid nature. Although hybrid applications can use the native functionalities of a device, they are still a little slower than native apps. Thus the method would be of better applicability if the DTN is a native app.[12][23]

Besides this, rather than storing the data in DAT file, the accelerometer values can be sent as CSV lines

## Author Biography

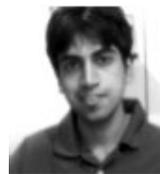

.**Rahul R Upadhyay** is a B.Tech student in Department of Mechanical engineering, BBD National Institute of Technology and Management, India. He has an interdisciplinary research interest in topics varying from embedded electronics, network security, and Mechatronics. He has published papers in national as well as international journals and participated and won several technical awards in national level competitions and conferences.